\documentclass[%
longbibliography,
 reprint,
superscriptaddress,
 amsmath,amssymb,
 aps,
prb,
floatfix,
]{revtex4-2}
\usepackage[english]{babel}
\usepackage{comment}
\usepackage{amsmath}
\usepackage{natbib}
\usepackage{ulem}
\usepackage{graphicx}
\usepackage{dcolumn}
\usepackage{bm}
\usepackage{amsmath}
\usepackage{siunitx}
\usepackage{listings}
\usepackage{mathtools}
\usepackage{float}
\usepackage{multirow}
\usepackage{pifont}
\usepackage{booktabs}

\usepackage{xcolor}
\usepackage{url}
\usepackage{hyperref}
\hypersetup{
    colorlinks=true,allcolors=blue,}
\begin{document}
\preprint{APS/123-QED}

\title{Field-free Josephson diode effect in a $d$-wave superconductor heterostructure}
\author{Hamed Vakili}
\affiliation{Department of Physics and Astronomy and Nebraska Center for Materials and Nanoscience, University of Nebraska, Lincoln, Nebraska 68588, USA}
\author{Moaz Ali}
\affiliation{Department of Physics and Astronomy and Nebraska Center for Materials and Nanoscience, University of Nebraska, Lincoln, Nebraska 68588, USA}
\author{Alexey A. Kovalev}
\affiliation{Department of Physics and Astronomy and Nebraska Center for Materials and Nanoscience, University of Nebraska, Lincoln, Nebraska 68588, USA}
\date{\today}

\begin{abstract}
We study superconductor/normal region/superconductor (S$|$N$|$S) Josephson junction formed using superconductors with $d$, $d+id^\prime$, and $d+is$ superconducting pairings. We show that the quality factor of the Josephson diode effect and its sign can be substantially tuned by the external magnetic field, gate voltage, and the length of the junction for all three types of pairings. We also identify the conditions under which the anomalous Josephson and Josephson diode effects can appear in the junction by analyzing appropriate symmetries. In particular, by breaking  a $\pi$-rotation symmetry, we show how a large field-free Josephson diode effect can be realized even in the absence of spin-orbit coupling. We also study the role of edge states appearing in the case of chiral superconductor with $d+id^\prime$ pairing. Our results demonstrate that the Josephson diode effect in a planar geometry can be used as a signature of unconventional superconducting pairings.
\end{abstract}
\pacs{Valid PACS appear here}
\maketitle 
\section{Introduction}
The superconducting diode (SDE) and Josephson diode effects (JDE) have attracted considerable attention in recent years due to many potential applications in superconducting devices and quantum computing~\cite{JDEexp,Miyasaka2021,Zhang2020,Wakatsuki2017,Lyu2021,Narita2022,Strambini2022,Theoryphenom, Theoryofdwave, fieldfreevander,JDEreview,Bocquillon2016,Pal2022,Baumgartner2022,Gupta2023}. The effects are characterized by non-reciprocal behavior in the critical current, in particular, the current can be dissipationless in one direction and dissipative in the opposite direction. In the absence of both time-reversal and inversion symmetry, Josephson junctions (JJ) can exhibit the JDE, characterized by non-reciprocal behavior in the critical supercurrent ($|I_c|$)~\cite{PhysRevX.12.041013,PhysRevB.103.245302,PhysRevB.103.144520,He2022,intrinsicdiode,Yuan2022,PhysRevLett.128.177001,PhysRevB.105.104508,PhysRevB.106.224509,PhysRevLett.129.267702,PhysRevB.106.134514}. This means that the critical supercurrent in the “$+$” direction ($|I^+_c|$) differs from that in the opposite “$-$” direction ($|I^-_c|$). This nonreciprocity in supercurrents holds promise for applications in superconducting electronics~\cite{Braginski2018,JJapp}.

The $d$-wave superconductivity is very common and appears in cuprate superconductors. The bulk of traditional $d$-wave superconductors is gapless. Recently, chiral $d$-wave superconductivity has been proposed in magic angle bilayer twisted graphene~\cite{twistedblgd+id,magicgraphene}, in Bi/Ni bilayers~\cite{Gong2017}, and in  Sn/Si(111) \cite{Gong2017}. It has been predicted that twisted bilayer cuprates can realize $d+is$ or $d+id^\prime$ pairings without gapless states in the bulk~\cite{Sigrist1998,twistedcuprate}; however, further research is necessary to confirm these predictions~\cite{expJJcuprate,Wang2023}. With the $d+id^\prime$ pairing, the bulk states are gapped, and instead, chiral edge states appear. Proximity induced superconductivity has been suggested in two-dimensional electron gas (2DEG) materials such as Bi$_2$O$_2$Se proximized with cuprates~\cite{PhysRevB.106.205424}. Furthermore, it has been demonstrated that a twisted bilayer cuprate exhibits time-reversal symmetry (TRS) breaking and JDE at specific twist angles~\cite{JJtwistedcuprate, twistedtj,expJJcuprate}.

\begin{figure}
    \centering
    \includegraphics[width=\columnwidth]{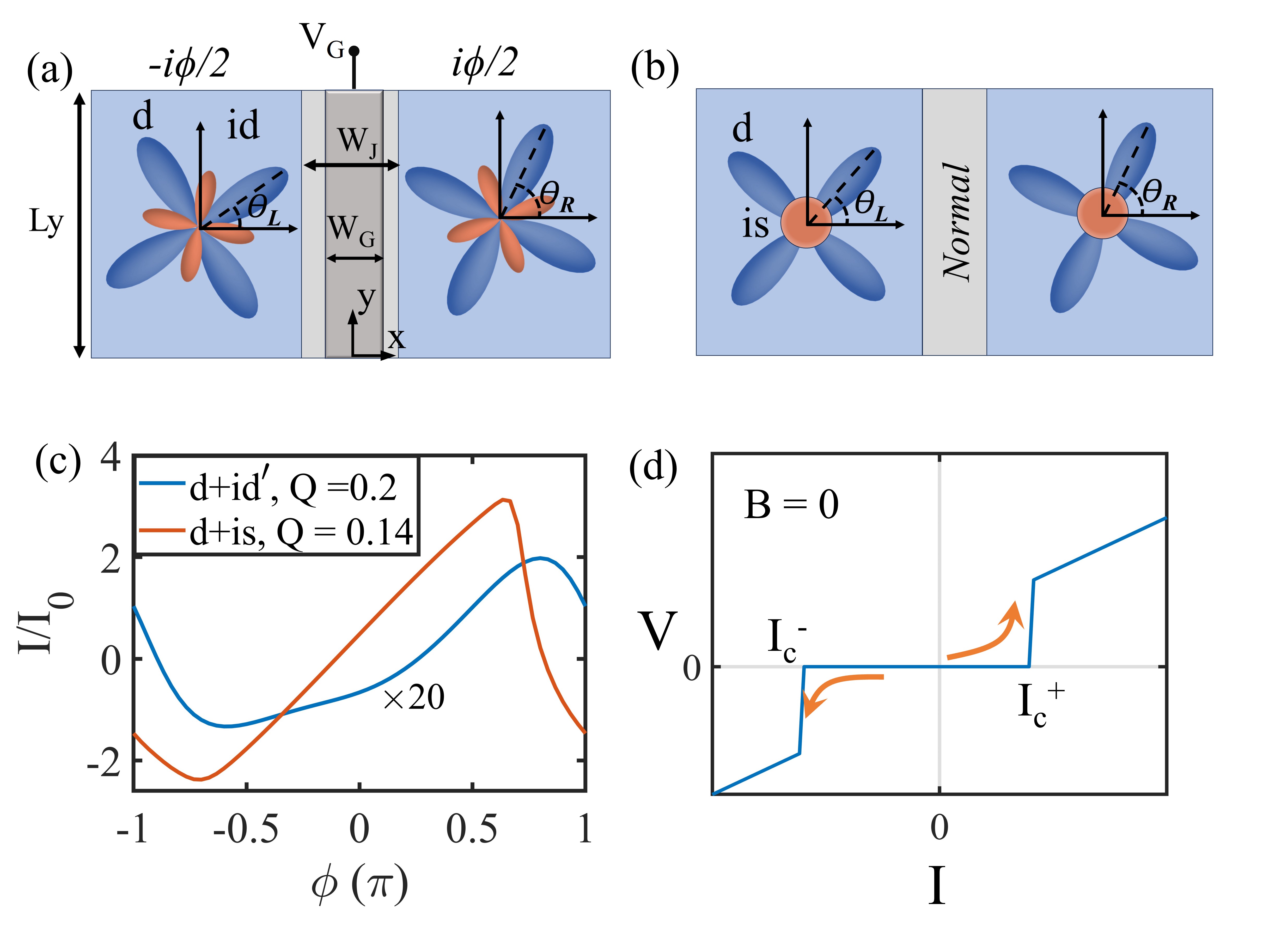}
    \caption{(a), (b) The schematics of the JJ used and the pairing lobes shown for $d+id^\prime$ and $d+is$ superconductor pairing in the leads, respectively. The shaded area of the normal region is the gated portion. For both $d+id^\prime$ and $d+is$ pairing, the left and right planes are rotated so that the pairing lobes have different angles, which breaks the necessary symmetries. (c) parameters used are: $\Delta_0 = 0.01t$, $\Delta_0^\prime = 0.001t$, $k_BT = 0.01 \Delta_0$, $L_y =20\xi ~(d+id^\prime), 10\xi ~(d+is)$, $W_J = \xi/2$, $\mu =0.1 ~(d+id^\prime), 0.2 t~(d+is)$. $I_0=e\Delta_0L_y /\hbar \xi$. $\theta_L = -\theta_R = 0.15 \pi$ for $d+id^\prime$ and $\theta_L = 0 \pi, \theta_R = 0.5 \pi$ for $d+is$ pairing. (d) The schematic of I-V plot for the JJ. Upward (downward) sweep measurement determines $I_c^+$ ($I_c^-$).    }
    \label{fig:d-wave}
\end{figure}

In this work, we study superconductor/normal region/superconductor (S$|$N$|$S) Josephson junction formed with $d$-wave superconductors. We consider $d$, $d+id^\prime$, and $d+is$ superconducting pairings that have been suggested in exfoliated layers of cuprate superconductors. We find that for the case of $d$-wave pairing, the finite JDE can only occur in the presence of spin-orbit coupling (SOC) and magnetic field.
The  $d+id^\prime$ and $d+is$ pairings break TRS. As a result, in the presence of additional asymmetry (see Fig.~\ref{fig:d-wave}) the TRS breaking results in JDE even in the absence of SOC and magnetic field. According to our results for $d+id^\prime$ pairing, the chiral edge modes show a clear contribution to JDE, which suggests that interference effects can be used to control JDE. We further confirm this by studying the distribution of the Josephson current in the junction and by identifying the Andreev bound states (ABS) from the local density of states (LDOS). Our proposal realizes a large, field-free JDE and demonstrates that JDE in a planar geometry can be used as a signature of unconventional superconducting pairings.

\section{Model and methods} 
We consider S$|$N$|$S Josephson junction (JJ) formed
using 2DEG with induced $d$-wave, $d+id^\prime$, or $d+is$ pairing, see Fig.~\ref{fig:d-wave}. An electrical gate $V_g$ is applied to the normal region as shown in Fig.~\ref{fig:d-wave}, which allows for the tunability of JJ. 

The Bogoliubov–de Gennes (BdG) Hamiltonian of the system, written in the Nambu basis ($\psi_\uparrow,\psi_\downarrow,\psi^\dagger_\downarrow,-\psi^\dagger_\uparrow$) in the continuous limit reads~\cite{Majdwave}
\begin{align}
H &=\left(\frac{\hbar^2\boldsymbol{k}^2}{2m^*} -(\mu+V)+\alpha_{SO}\left(\hat{\boldsymbol{\sigma}} \times \boldsymbol{k}\right)_z \right) \hat{\tau}_{z}\nonumber\\  
&+ \boldsymbol{h}.\boldsymbol{\sigma}+\hat{\tau}_x \text{Re}\Delta(\boldsymbol k,x) -\hat{\tau}_y\text{Im}\Delta(\boldsymbol k,x).\label{eq:Hamiltonian}
\end{align}
Here, $m^*$ is the effective electron mass, $\mu$ is the chemical potential, $\alpha$ is the Rashba spin-orbit coupling strength, and $\boldsymbol{h}$ describes the Zeeman energy, e.g., for an external magnetic field. The gate voltage is defined as $V= V_G\;\Theta(W_G/2-|x|)$. 
The pair potential $\Delta(\boldsymbol k,x)$ is parametrized for $d+id^\prime$ pairing as
\begin{align}\label{eq:did}
    \Delta(\boldsymbol k,x) &=  2e^{i\phi(x)}\Bigl[ \Tilde{\Delta}(x)[\cos{2\theta}\left(\cos{k_x}-\cos{k_y}\right)\nonumber \\ &+\sin{2\theta}\sin{k_x}\sin{k_y}]+\nonumber  \\& i\Tilde{\Delta}^\prime(x)[\cos{(2\theta+\pi/2)}\left(\cos{k_x}-\cos{k_y}\right)\nonumber \\ &+\sin{(2\theta+\pi/2)}\sin{k_x}\sin{k_y}]\Bigr],
\end{align}
and for $d+is$ pairing as
\begin{align}\label{eq:dis}
    \Delta(\boldsymbol k,x) &=  2e^{i\phi(x)}\Bigl[ \Tilde{\Delta}(x)[\cos{2\theta}\left(\cos{k_x}-\cos{k_y}\right)\nonumber \\ &+\sin{2\theta}\sin{k_x}\sin{k_y}]
    -i\Tilde{\Delta}^\prime[\cos{k_x}+\cos{k_y}]\Bigr],
\end{align}
where $\theta$ describes the angle between the $x$-axis and the lobe direction
of the pair potential.  We denote $\theta_L$ ($\theta_R$) for the left (right) superconductor.
The terms $\Tilde{\Delta}(x),\Tilde{\Delta}^\prime(x)$ are only nonzero in the regions covered by superconductors where $\Tilde{\Delta}(x)=\Delta_0$, $\Tilde{\Delta}^\prime(x)=\Delta^\prime_0$. For the region covered by the left superconductor, we take $\phi(x)=\phi_L$, and for the region covered by the right superconductor, we take $\phi(x)=\phi_R$. We discretize the Hamiltonian \eqref{eq:Hamiltonian} and use the tight binding approach where $t=\frac{\hbar^2}{2 m^* a^2}$ is the nearest neighbor hopping and $a$ is the lattice spacing. We use dimensionless units measuring energy in units of $t$ and length in units of $a$. The SC coherence length is defined as $\xi =\frac{\hbar^2 k_F}{m^* \Delta_0} =2a{\frac{\sqrt{\mu t}}{\Delta_0}}$. Typical material parameters we use in numerical calculations are as follows: $a = 2.5 nm$, $m^* = 0.14 m_e$, $\alpha_{SO}=0.1ta=0.1 eV {\textup{\AA}}$, $\Delta_0 = 0.01t=0.4 meV$, and $\mu =0.1t= 4 meV$~\cite{Wu2017}.

We calculate the conserved current inside the junction (see Fig.~\ref{fig:currentdensity}(a)) which does not require the self-consistent calculation.
The self-consistent procedure would be required inside superconductor to ensure the current conservation. 
To calculate the equilibrium Josephson current, we use the Matsubara Green's function $G(i\omega_n)$ obtained using the analytical continuation of the retarded Green's function. The local current between the sites $\mathbf{n}$ and $\mathbf{m}$ can be expressed using the standard approach~\cite{Asano,prlJJ,PhysRevB.94.094514,PhysRevLett.119.187704,PhysRevLett.120.047702,Baumgartner2021}:
\begin{eqnarray} 
I_{\mathbf{n},\mathbf{m}} = 2\frac{ek_BT}{\hbar} \sum_\mathbf{\omega_\mathbf{n}}\mathrm{Tr}\left[Im\left(H_{\mathbf{n}\mathbf{m}}G_{\mathbf{m}\mathbf{n}}-H_{\mathbf{m}\mathbf{n}}G_{\mathbf{n}\mathbf{m}}\right)\right],
\end{eqnarray} 
where $G_{\mathbf{n}\mathbf{m}}$ is the Green's function and $H_{\mathbf{n}\mathbf{m}}$ is the Hamiltonian submatrix calculated between the sites $\mathbf{n}$ and $\mathbf{m}$, with the lattice sites taken inside the normal region.
The summation is performed over the fermionic Matsubara frequencies $\omega_n = (2n+1)\pi k_B T$ and the Green's functions $G_{\mathbf{n}\mathbf{m}}$ also accounts for the self energy of the leads. The total current is obtained by summing all contributions along the cut in 
$y$-direction, 
$I=\sum_\mathbf{n}I_{\mathbf{n},\mathbf{n}+\mathbf{e}_x}$, where 
$\mathbf{e}_x$ is the unit vector along the 
$x$-axis (see Fig.~\ref{fig:currentdensity}(a)). The diode effect is then characterized by the quality factor $Q = (I^+_c-|I^-_c|)/(I^+_c+|I^-_c|)$, where $I^+_c$ and $I^-_c$ are the critical current calculated by finding the maximum and minimum of current over $\phi=\phi_L-\phi_R \in \{0, 2\pi\}$. As an example, in Fig.~\ref{fig:d-wave}(c) we show the current-phase relation (CPR) for $d+id^\prime$ and $d+is$ pairings. The figure shows the nonvanishing anomalous Josephson effect at $\phi=0$ provided that the two pairing lobes are rotated with respect to each other. Furthermore, CPR in the figure results in the quality factor $Q$ equal to $0.2$ for $d+id^\prime$ pairing and $0.14$ for $d+is$ pairing. 

We also calculate the local density of states (LDOS) using the expression:
\begin{align}
    \rho(\omega,n) = -\frac{1}{\pi}\mathrm{Tr_n}(\mathrm{Im}[G(\omega+i\eta)]),
\end{align}
where $i\eta$ is the band broadening. The location of LDOS peaks allows us to determine energies of the Andreev bound states (ABS).

\begin{table}[!h]
\begin{tabular}{c||c|c|c|l}
 \toprule
 SC Pairing& $\mathcal{R}_x(\pi)$&$\mathcal{R}_y(\pi)$&$\mathcal{T}$ &Symmetries to break or  \\
 & & & &terms to add for JDE\\
 \hline
 $d_{x^2-y^2}$&   $\checkmark $& $\checkmark $&$\checkmark$& $h \hat{\sigma}_y$ + Rashba SOC\\
 $d_{xy}$&   $\checkmark ^\star$& $\checkmark ^\star$&$\checkmark$&$h \hat{\sigma}_y$ + Rashba SOC\\
 $d_{x^2-y^2} + d_{xy}$& $\times$&$\times$&   $\checkmark$&$h \hat{\sigma}_y$ + Rashba SOC\\
 $d_{x^2-y^2} + id_{xy}$&$\times$& $\times$&  $\times$&$\mathcal{R}_x(\pi)\mathcal{R}_y(\pi)\& \mathcal{R}_x(\pi)\mathcal{T}$\\
 $d_{x^2-y^2} + is$&$\checkmark$& $\checkmark$&  $\times$ &$\mathcal{R}_y(\pi)$\\
\bottomrule
\end{tabular}
\caption{The symmetry of each SC pairing type and corresponding JJ needed to have JDE.$^\star$ With a $\pi$ gauge transformation of SC pairing.}
\label{table}
\end{table}
\begin{figure}
    \centering
    \includegraphics[width=\columnwidth]{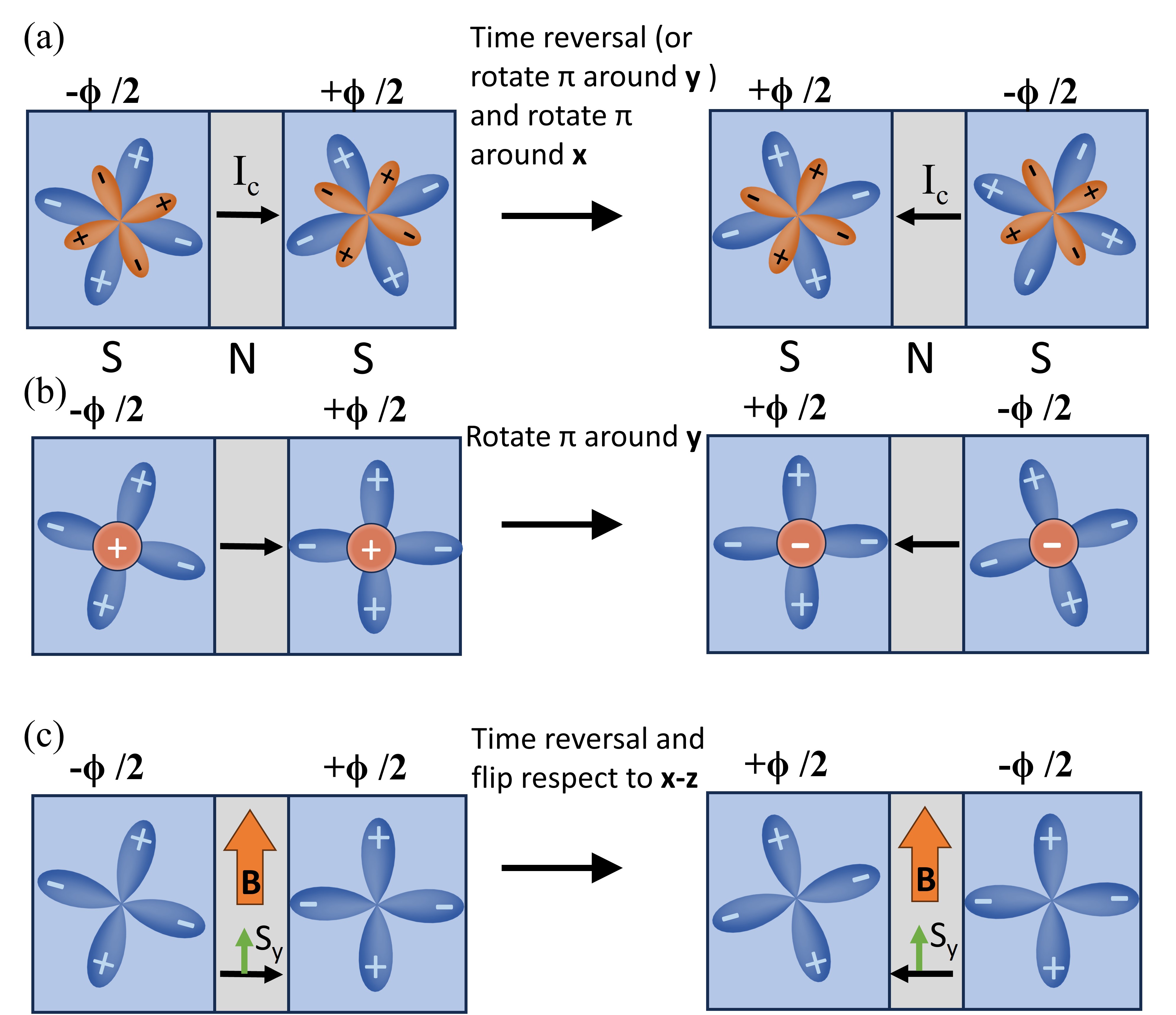}
    \caption{(a) The pictorial symmetry analysis of the JJ with $d+id$ pairing in the leads. $I_c$ is the current inside the normal region, shown by the black arrow. To recover the correct chirality and phase sign, a rotation around y-axis followed by a rotation around x-axis is needed (inversion with respect to the center of JJ). (b) With $d+is$ pairing in the leads, $\mathcal{R}_x(\pi)$ would recover $I(\phi) = -I(-\phi)$ only if left and right side satisfy $\beta_L = \beta_R$. (c) For $d_{x^2-y^2}$ pairing in the leads, we need non-zero magnetic and SOC terms. The SOC term prevents the spin rotation to recover the correct spin-momentum coupling.    }
    \label{fig:pictorial}
\end{figure}
\section{Symmetry analysis}  
To understand how the anomalous Josephson effect and JDE can be achieved, we begin with a symmetry analysis of the system, followed by numerical simulations of the diode effect while varying various parameters.
In order to get a non-reciprocal transport in a Josephson junction, it is required to have $I(\phi) \neq -I(-\phi) $. If however there is a symmetry $\Tilde{T}$ that reverses $I$ and $\phi$ then the anomalous Josephson effect and JDE (by extension of the above argument to applied voltage $V$) vanish. To identify such a symmetry, we start with analysis of systems with $d$-wave pairings. In Table~\ref{table}, we list whether TRS $\mathcal{T}$ and $\pi$-rotation symmetries with respect to $x$ and $y$ axis ($\mathcal{R}_x(\pi), \mathcal{R}_y(\pi)$) are satisfied for the system in Fig.~\ref{fig:d-wave} for simple alignment of the pairing lobes. In the last column, we list modifications that are necessary to break the symmetry or symmetries prohibiting the anomalous Josephson effect. For example, in case of $d_{x^2-y^2}+d_{xy}$ pairing, TRS is present, and $\mathcal{T}e^{i\phi}\Delta \mathcal{T}^{-1}=e^{-i\phi}\Delta $. This means that to achieve the nonreciprocity, we need to break both TRS and spin rotational symmetry. The choice of the direction of the magnetic field and the presence of Rashba SOC guarantee that we cannot use $\mathcal{R}_x(\pi)$ and $\mathcal{R}_y(\pi)$ symmetries for the transformation that reverses both $I$ and $\phi$. With $d_{x^2-y^2}+id_{xy}$ pairing, all of the three symmetries ($\mathcal{T}$, $\mathcal{R}_x(\pi)$, $\mathcal{R}_y(\pi)$) are separately broken. However, the combined symmetries $\mathcal{R}_x(\pi)\mathcal{R}_y(\pi)$ and $\mathcal{R}_x(\pi)\mathcal{T}$ still need to be broken for the presence of the anomalous Josephson effect. This can be achieved by using JJ with an asymmetry (e.g. induced by gating) or by rotating the pairing lobes of the left and right superconductors with respect to each other. Finally, for the $d_{x^2-y^2}+is$ pairing, the rotational symmetries are present. Since $\mathcal{R}_y(\pi)$ reverses both $I$ and $\phi$ it has to be broken for the presence of the anomalous Josephson effect. As before, this can be achieved by using JJ with an asymmetry (e.g. induced by gating) or by rotating the pairing lobes of the left and right superconductors with respect to each other. The latter approach is explored in details in our numerical simulations. 
Figure~\ref{fig:pictorial} shows the pictorial demonstration of the symmetry analysis. In particular, Fig.~\ref{fig:pictorial}(a) shows that for $d+id^\prime$ pairing the orientations of the pairing lobes in the left and right SCs (i.e. $\theta_L\neq\theta_R$) can break $\mathcal{R}_x(\pi)\mathcal{R}_y(\pi)$ and $\mathcal{R}_x(\pi)\mathcal{T}$ symmetries. In Fig.~\ref{fig:pictorial}(b), we show that for $d+is$ pairing, the orientations of the pairing lobes (i.e. $\theta_L\neq\theta_R$) break $\mathcal{R}_y$ symmetry provided that $\theta_L\neq -\theta_R$. In Fig.~\ref{fig:pictorial}(c), we assume $d_{x^2-y^2} + d_{xy}$ pairing. Here, a Zeeman field along the $y$ axis is necessary to break TRS. To further break the spin rotation symmetries, the Rashba SOC term has to be added. It is worth mentioning that the anomalous Josephson effect has been predicted for $s$-wave pairing without the presence of SOC by introducing appropriate magnetic textures~\cite{PhysRevB.108.214520}.

\section{Leads with $d$-wave pairing}  
Here, we consider a ballistic JJ with superconducting leads featuring the pure $d$-wave pairing. To treat systems with large $L_y$, we assume here periodic boundaries along the $y$ axis and numerically find solutions characterized by the transverse wave vector $k_y$. As a result, we solve numerically the effective one-dimensional scattering problem along the direction of the junction, afterwards summing the contributions from each $k_y$.  With this approach, we are able to reproduce analytical results in Ref.~\cite{Du2023} where the Zeeman and Rashba terms were only present in the junction region. 

\begin{figure}
    \centering
    \includegraphics[width=\columnwidth]{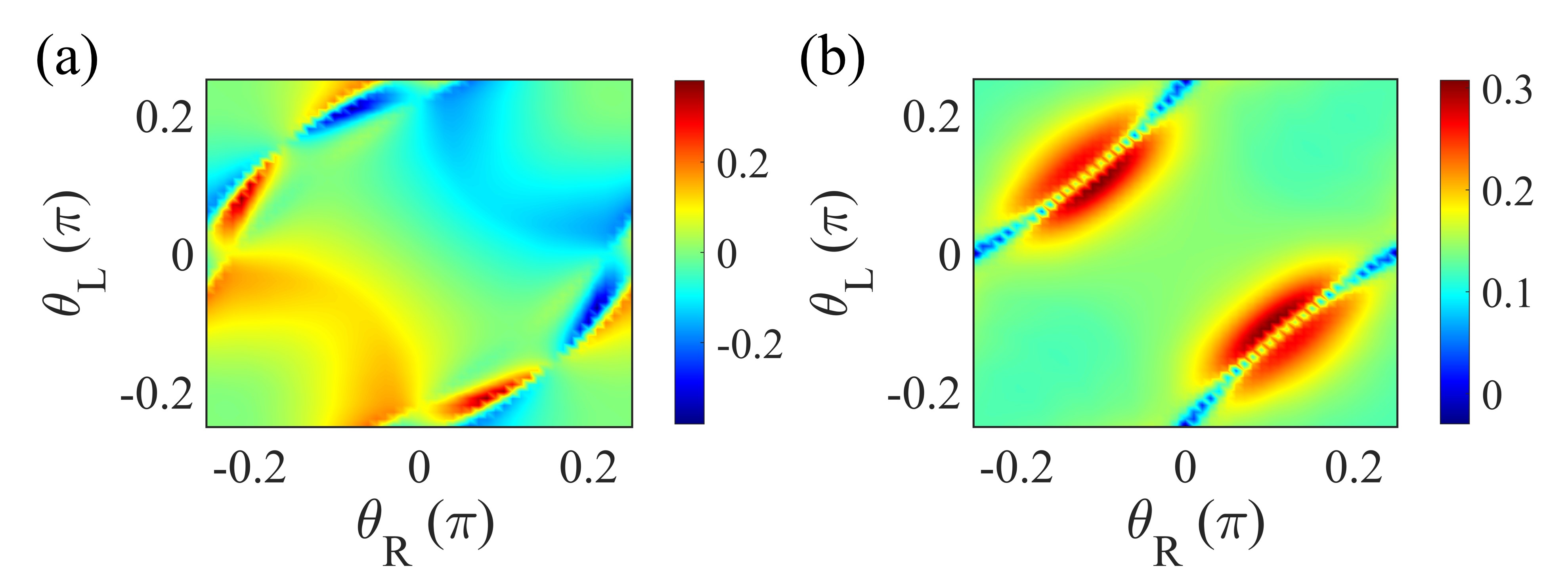}
    \caption{Q is plotted as $\theta_L$ and $\theta_R$ is varied. The Zeeman energy is in the x, y direction for (a), (b), respectively. Parameters used: $\mu = 0.1\:t$, $\alpha_{SO} = 0.1\:ta$, $\Delta_0 = 0.01\:t$, $\Delta^\prime_0 = 0\:t$, $k_BT = 0.01\:\Delta_0$, $W_J = \xi/2$, $|h| = 0.1\:\Delta_0$. (a) $h=(h_x,0,0)$. At the off diagonal values ($\theta_L=-\theta_R$) Q is zero. (b)  $h=(0,h_y,0)$. The calculations are performed using a Fourier transform along the y-axis.   }
    \label{fig:dwave}
\end{figure}

For an arbitrary orientation of the pairing lobes, the pairing potential can take a general form of $d_{x^2-y^2}+d_{xy}$. For nonvanishing JDE, as shown before, we need the Zeeman term to break the time-reversal symmetry, and the Rashba SOC to break the spin rotation symmetry. Figure~\ref{fig:dwave} shows the quality factor $Q$ plotted as a function of the pairing lobe orientation angles $\theta_L$ and $\theta_R$. Variations in the lobe orientation can be achieved by adjusting the crystal orientation of SC leads. In Fig.~\ref{fig:dwave}(a), the Zeeman field is applied in the $x$ direction. The figure shows regions of large $Q$ throughout the phase space. In Fig.~\ref{fig:dwave}(b), the Zeeman field is applied in the $y$ direction. For this direction of the Zeeman field, even with no rotation of lobes, i.e., for $\theta_L=0$ and $\theta_R=0$, we observe large values of $Q$. Finally, we  observe relatively weak JDE for the Zeeman field along the $z$ axis (not shown in the figure) in contrast to a JJ based on the topological insulator~\cite{Theoryofdwave}.

\begin{figure}
    \centering
    \includegraphics[width=\columnwidth]{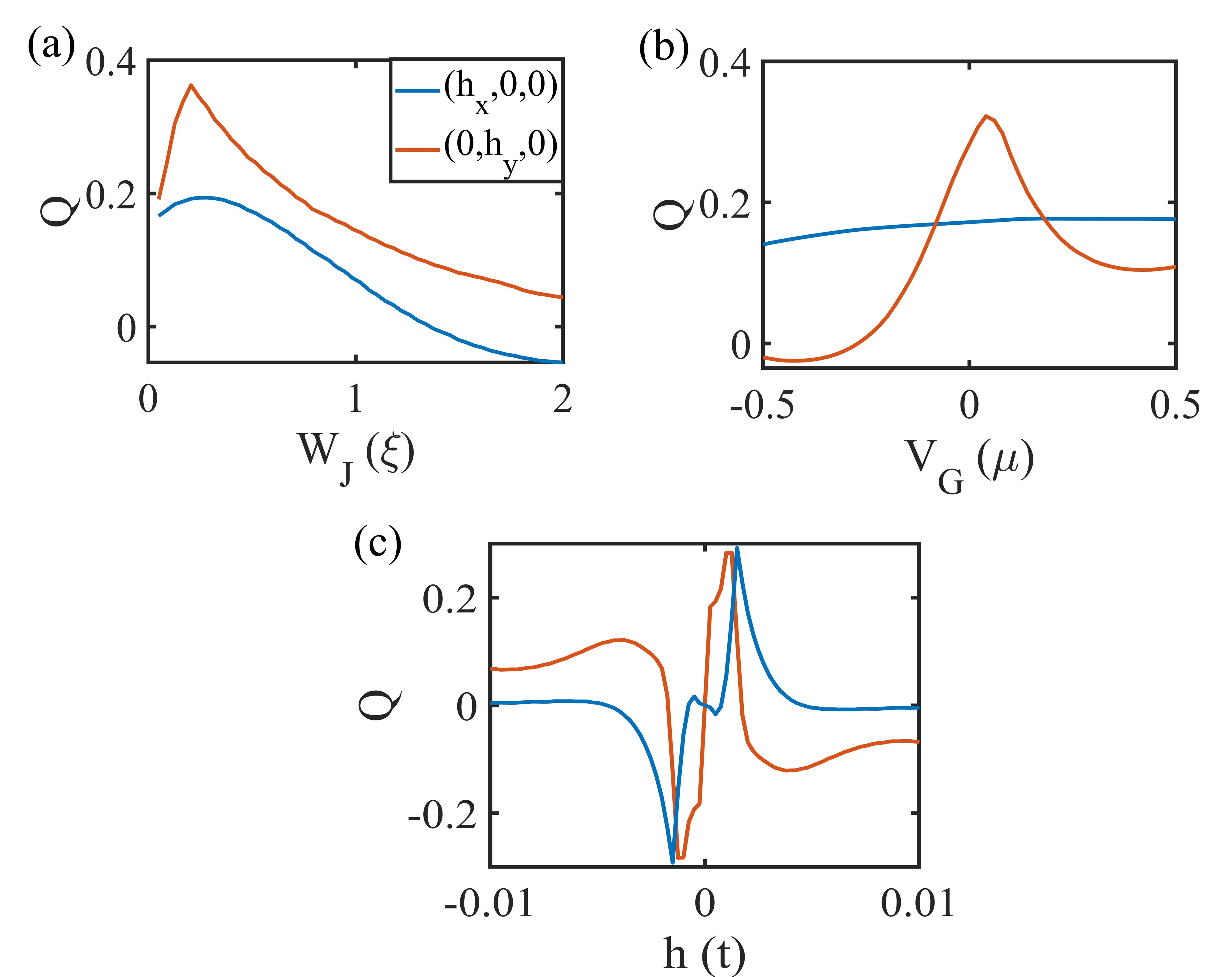}
    \caption{(a) The plot of Q as $W_J$ is increased for two cases of Zeeman energy in the x and y directions. (b) The plot of Q as the gate potential $V_G$ is varied. Parameters used: $\mu = 0.1\:t$, $\alpha_{SO}=0.1\:ta$, $\Delta_0 = 0.01\:t$, $\Delta^\prime_0 = 0$, $k_BT = 0.01\:\Delta_0$, $\theta_L = \theta_R = -0.15\:\pi$ for $h = (h_x,0,0)$ and  $\theta_L = -\theta_R = -0.10\:\pi$ for $h = (0,h_y,0)$ plots. $h_x = 0.2\:\Delta_0$, $h_y = 0.1\:\Delta_0$ is used for (a) and (b) plots. $W_J = \xi/2$ in (b) and (c) plots. $W_G = \xi/8$ in (b). (c) The plot of Q as $|h|$ is varied in $x$ and $y$ directions.}
    \label{fig:Q_gate_wj}
\end{figure}
\begin{figure}
    \centering
    \includegraphics[width=\columnwidth]{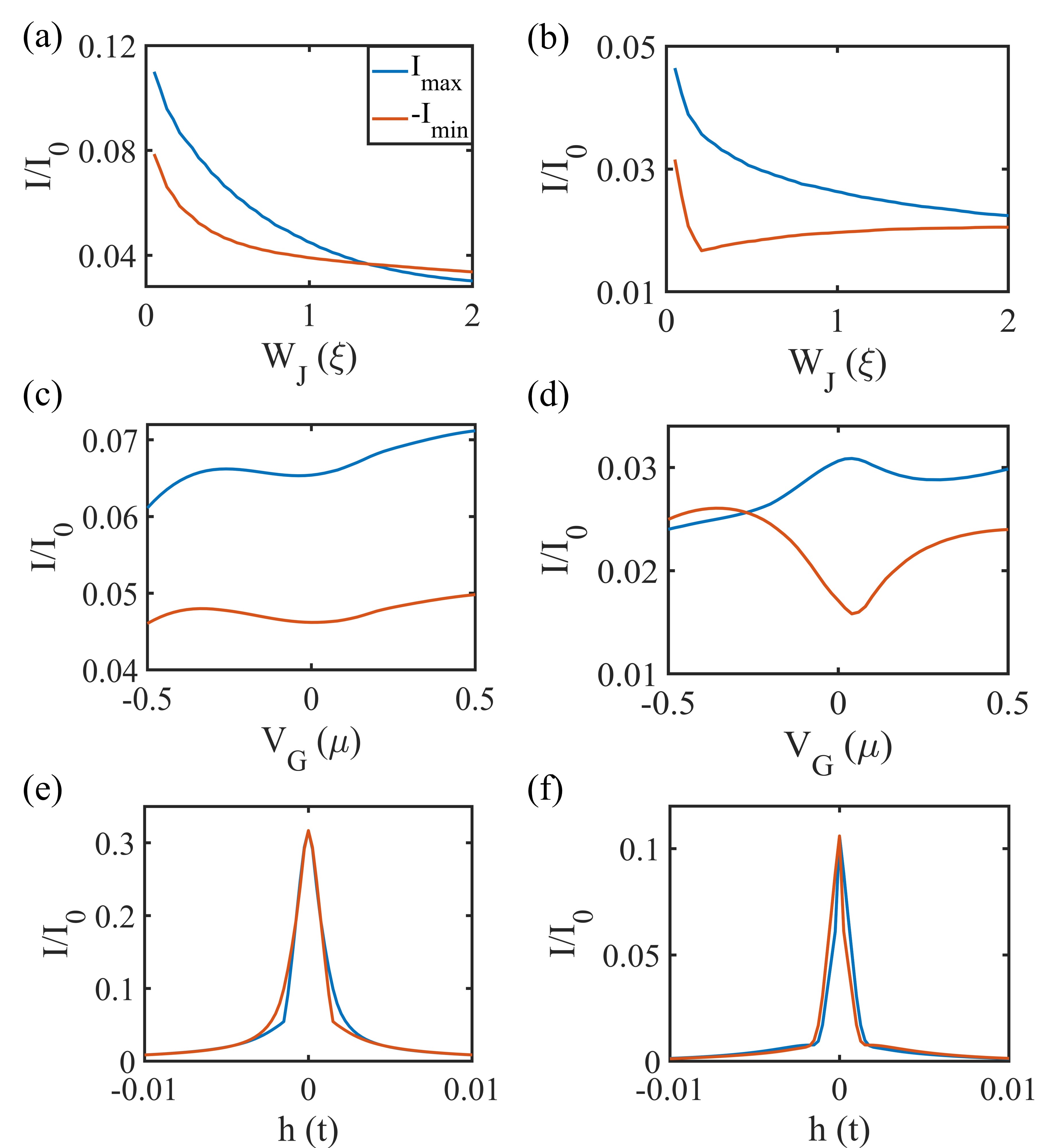}
    \caption{$I_{max}$ and $I_{min}$ are plotted as $W_J$ is varied for h in $x$ and $y$ direction in (a) and (b), respectively. (e) and (f) are plots of $I_{max}$ and $I_{min}$ as $|h|$ is varied. The parameters used are the same as Fig.~\ref{fig:Q_gate_wj}.}
    \label{fig:Imaxmin}
\end{figure}

Several parameters can be adjusted to fine-tune the quality factor of JJ. Fist, we vary the junction width $W_J$. As shown in Fig.~\ref{fig:Q_gate_wj}(a), the quality factor shows the large variation and even sign change as $W_J$ is increased.  
This behavior stems from contributions of the higher order harmonics to the current-phase relation (CPR). 
Next, we also explore how $Q$ depends on the gate voltage $V_G$ and the Zeeman field $\mathbf h$. Figure~\ref{fig:Q_gate_wj}(b) demonstrates that $Q$ can be effectively tuned by the gate voltage within the junction region for the Zeeman field along the $y$ axis. The variation of $Q$ as a function of the Zeeman field along the $x$ and $y$ directions is shown in Fig.~\ref{fig:Q_gate_wj}(c). In Fig.~\ref{fig:Q_gate_wj}(c), we observe large variations as the Zeeman field is increased, with eventual suppression of JDE at large Zeeman fields. 
Even though the dependence of the quality factor $Q$ on various parameters is complicated, some trends can be identified. This can be useful in fine tuning the diode effect for applications such as Josephson transistor. 

This is further elucidated in Fig.~\ref{fig:Imaxmin} showing the evolution of the maximum, $I_{max}$, and minimum, $I_{min}$, Josephson currents. Figure~\ref{fig:Imaxmin} shows the evolution of $I_{max}$ and $I_{min}$ for the calculations in Fig.~\ref{fig:Q_gate_wj}, for the Zeeman field $h$ along the $x$ axis in (a), (c), and (e) plots, and for the Zeeman field $h$ along the $y$ axis in (b), (d), and (f) plots. The high tunability of Josephson current is seen in Fig.~\ref{fig:Imaxmin}(b) where a dip in $I_{min}$ corresponds to the peak observed in $Q$. In comparison, Fig.~\ref{fig:Imaxmin}(a) shows a monotonic decay in both $I_{max}$ and $I_{min}$ corresponding to the smoother change in $Q$, as can be seen in Fig.~\ref{fig:Q_gate_wj}(a). In the long junction limit $W_J \gg \xi$, we observe that $Q$ tends to zero. By looking at the variation of $I_{max}$ and $I_{min}$ as the gate voltage $V_G$ is varied in Fig.~\ref{fig:Imaxmin}(c), we observe only minor change. However, Fig.~\ref{fig:Imaxmin}(d) demonstrates high tunability by gate voltage as we observe a dip in $I_{min}$ and a peak in $I_{max}$ at around $V_G=0$. In this case, the gating also substantially modifies $Q$, even leading to a sign change. Figures~\ref{fig:Imaxmin}(e) and (f) show large $I_{max}$ and $I_{main}$ at small $h$. Away from the region $h = 0$ the magnetic field suppresses the supercurrent. We observe that the Zeeman field skews $I_{max}$ and $I_{min}$ in the opposite directions, which correspond to the sign change of $Q$ seen in Fig.~\ref{fig:Q_gate_wj}(c). Overall, Figs.~\ref{fig:Q_gate_wj} and \ref{fig:Imaxmin} demonstrate high tunability of JDE for leads with the $d$-wave pairing.

\section{Leads with $d \pm id^\prime$ pairing}  
\begin{figure}
    \centering
    \includegraphics[width=\columnwidth]{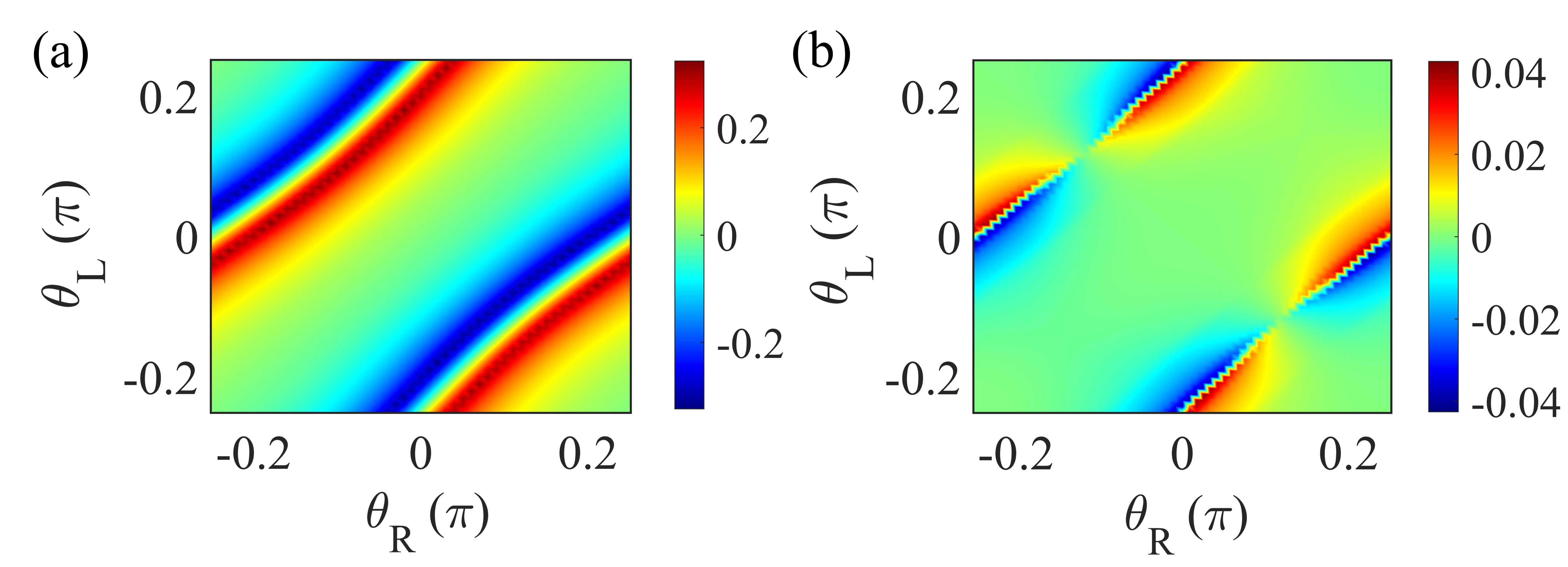}
\caption{Q phase space as $\theta_L$, $\theta_R$ are varied for a $d+id^\prime|$N$|d+id^\prime$ JJ in (a) and a $d+id^\prime|$N$|d+id^\prime$ JJ in (b). (a) Along the $\theta_R=\theta_L$ line, Q is zero. (b) In this case, on the off-diagonal line ($\theta_R=-\theta_L$), Q is zero. Parameters used: $\mu = 0.1\:t$, $\Delta_0 = 0.01\:t$, $\Delta^\prime_0 = 0.001\:t$, $k_BT = 0.01\:\Delta_0$,  and $W_J = \xi/2$. The calculations are performed using a Fourier transform along the y-axis. }
    \label{fig:Q_phase}
\end{figure}

\begin{figure}
    \centering
    \includegraphics[width=\columnwidth]{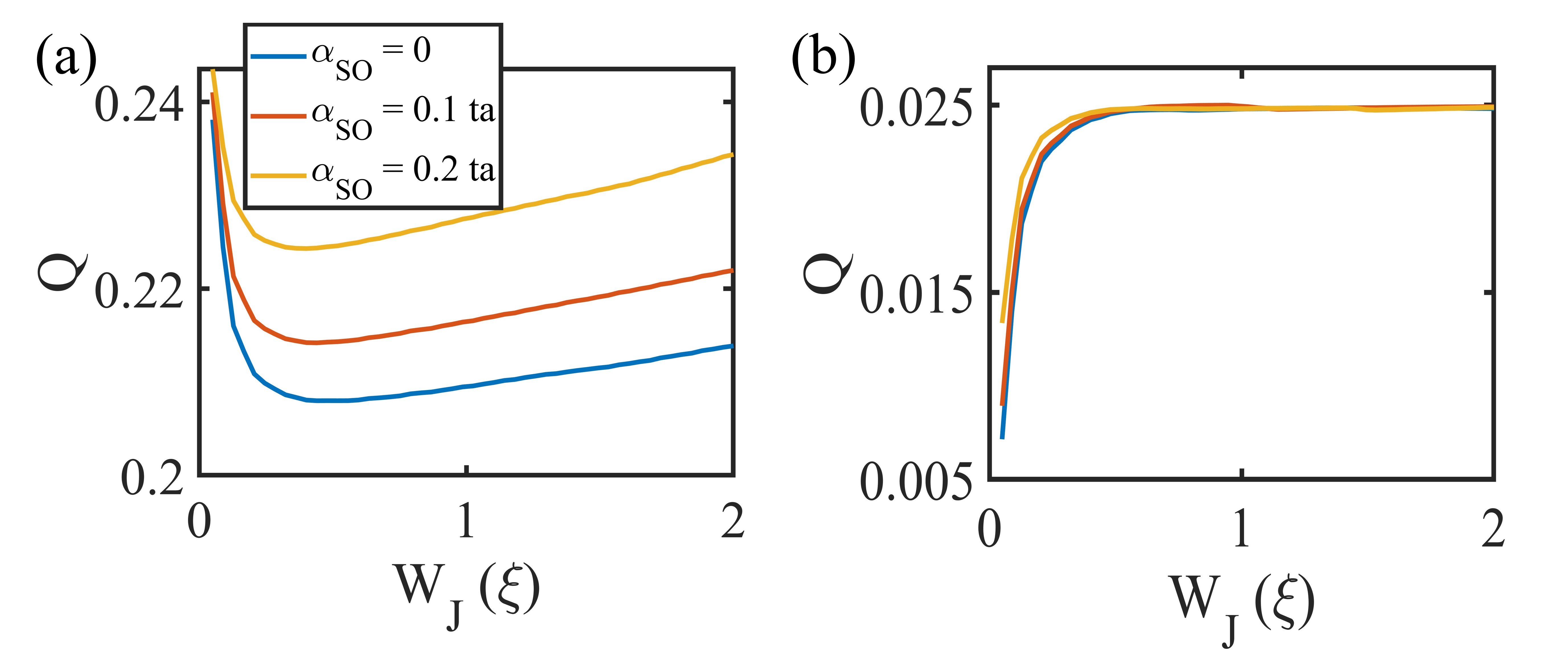}
\caption{The plot of Q as $W_J$ is increased for various spin-orbit coupling strength $\alpha_{SO}$ for the two cases of the same chirality ($d+id^\prime|$N$|d+id^\prime$) and opposite chirality ($d-id^\prime|$N$|d+id^\prime$) in (a) and (b), respectively. Parameters used: $\mu = 0.1\:t$, $\Delta_0 = 0.01\:t$, $\Delta^\prime_0 = 0.001 \:t$, $k_BT = 0.01 \:\Delta_0$.  $\theta_L = -\theta_R = -0.15\:\pi$ used for (a) and $\theta_L=0$, $\theta_R=0.2\:\pi$ used for (b).      }
    \label{fig:Q_geometry}
\end{figure}

Now we consider a ballistic JJ with superconducting leads featuring the $d\pm id^\prime$ pairing. For this pairing, the TRS is broken. To satisfy the conditions of JDE, we still need to break rotational symmetries (see Table~\ref{table}). This can be achieved by rotating the pairing lobes parametrized by angle $\theta$ in Eq.~\eqref{eq:did}.  Figure~\ref{fig:Q_phase}(a) shows the quality factor $Q$ as a function of $\theta_L$ and $\theta_R$ without any Zeeman field or SOC for a $d+id^\prime|$N$|d+id^\prime$ JJ. We observe a possibility of large JDE even in the absence of the Zeeman field and SOC. As the system may have the Rashba SOC, we also repeated the calculations with the inclusion of Rashba SOC, finding only minor differences. Figure~\ref{fig:Q_phase}(b) shows the quality factor $Q$ as a function of $\theta_L$ and $\theta_R$ without any Zeeman field or SOC for a $d-id^\prime|$N$|d+id^\prime$ JJ. Compared to the previous case, we observe significantly smaller values of $Q$. For this case no JDE is present on the off diagonal, i.e., $\theta_L=-\theta_R$. We further look at how Q evolves as $W_J$ is varied. Figure~\ref{fig:Q_geometry}(a) shows $Q$ for the case of a $d+id^\prime|$N$|d+id^\prime$ JJ. The short junction limit $W_J \ll \xi$ shows the largest $Q$, although the decrease in $Q$ as $W_J$ is increased is not significant. Figure~\ref{fig:Q_geometry}(b) is the calculations for a $d-id^\prime|$N$|d+id^\prime$ JJ. As before, we observe much smaller $Q$ for superconductors with opposite chiralities.

\begin{figure}
    \centering
    \includegraphics[width=\columnwidth]{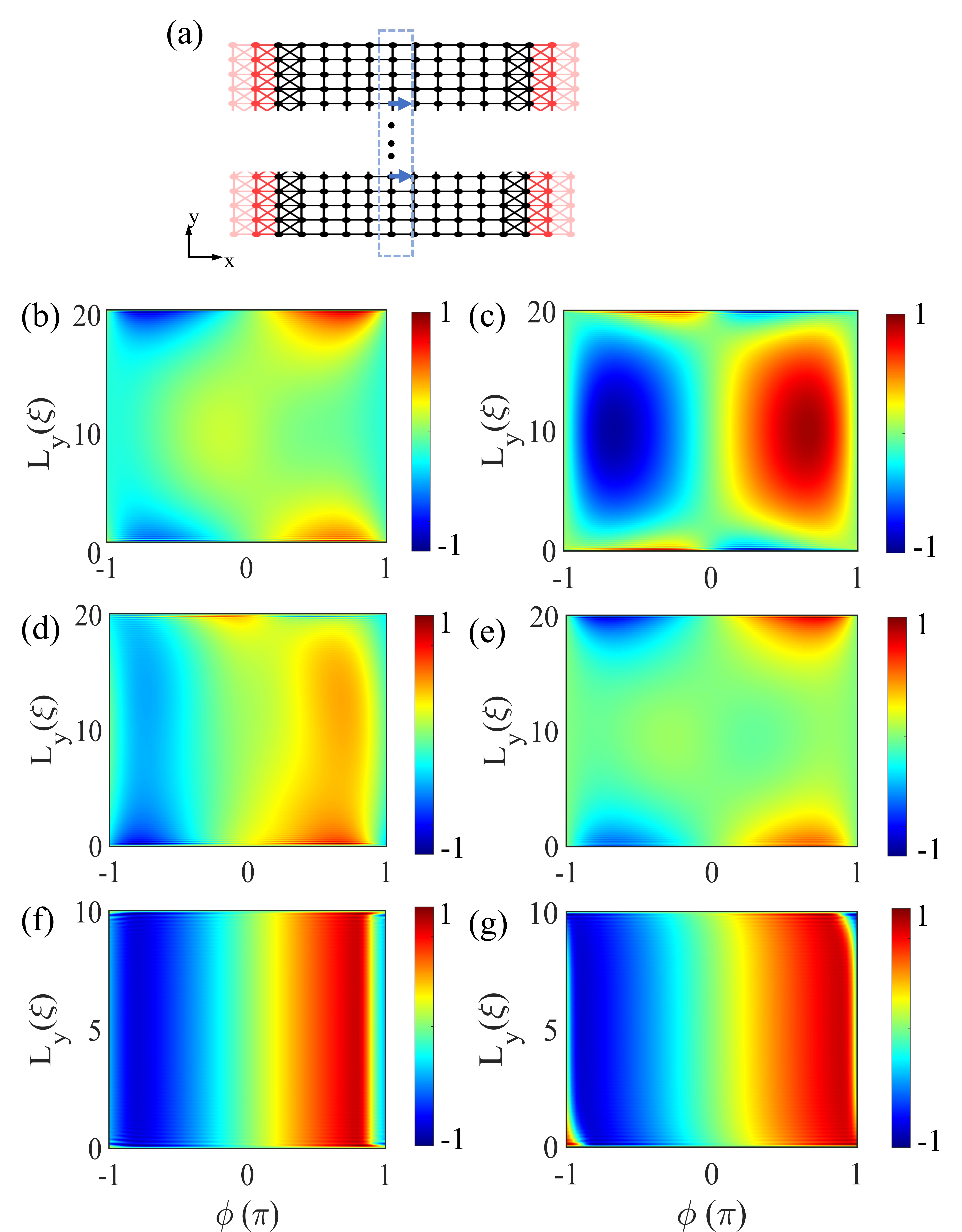}
    \caption{(a) Schematics of the tight binding of JJ with SC leads at the left and right sides. The current distribution along y as $\phi$ is varied for a $d+id^\prime|$N$|d+id^\prime$ JJ with $\theta_L = -\theta_R = 0.15\:\pi$ in (b) and $\theta_L = \theta_R = 0.15\:\pi$ in (c). The JJ in (d) and (e) is the $d-id^\prime|$N$|d+id^\prime$ type.  $\theta_L =0$, $\theta_R = 0.2\:\pi$ is used in (d) and $\theta_L = -\theta_R = 0.15\:\pi$ used in (e). Current density - phase relation with $d+is$ pairing shown in (f) with $\theta_L =0$,  $\theta_R = 0.5\:\pi$ and in (g) with $\theta_L =\theta_R = 0.15\:\pi$. The dark red of the colormap corresponds to the minimum of the current distribution, and the dark blue corresponds to the maximum of the current distribution in each plot. Parameters used: $\mu = 0.1\:t$, $\Delta_0 = 0.01\:t$, $\Delta^\prime_0 = 0.001\:t$,  $k_BT = 0.01\:\Delta_0$, $W_J = \xi/2$.  }
    \label{fig:currentdensity}
\end{figure}

In the case of $d+id^\prime$ pairing, the superconductors exhibit chiral edge states, with the relative sign of the $d$ and $d^\prime$ components determining the chirality of the system. To investigate how edge currents contribute to JDE, we examine the current distribution along the transverse direction for a JJ of finite width $L_y$. Figure~\ref{fig:currentdensity}(a) illustrates the schematics of the current distribution calculation. We consider a $d+id^\prime|$N$|d+id^\prime$ JJ in Figs.~\ref{fig:currentdensity}(b) and (c). We observe a concentration of Josephson current at the top and bottom edges, consistent with the effect of edge states. As expected, in Fig.~\ref{fig:currentdensity}(c) we observe vanishing JDE. In Fig.~\ref{fig:currentdensity}(b), the current in the bulk of the channel exhibits a significantly stronger anomalous Josephson effect compared to edges, and the edge part of the current mostly follows a $\sin{\phi}$ CPR behavior resulting in relatively weak contribution to JDE. Figures~\ref{fig:currentdensity}(d) and (e) show the current distribution for a $d-id^\prime|$N$|d+id^\prime$ JJ. In Fig.~\ref{fig:currentdensity}(e), we observe vanishing JDE. In Fig.~\ref{fig:currentdensity}(d), both the bulk and the edge modes seem to contribute to JDE; however, overall JDE is much weaker in this case. If we compare the current distribution calculations for $d\pm id^\prime$ in Figs.~\ref{fig:currentdensity}(b--e) with similar calculations for $d\pm is$ case in Figs.~\ref{fig:currentdensity}(f, g), we observe a clear difference that can be associated with the edge states present in $d\pm id^\prime$ systems. Figures~\ref{fig:currentdensity}(f, g) will be discussed in more detail in the next section.

\begin{figure}
    \centering
    \includegraphics[width=\columnwidth]{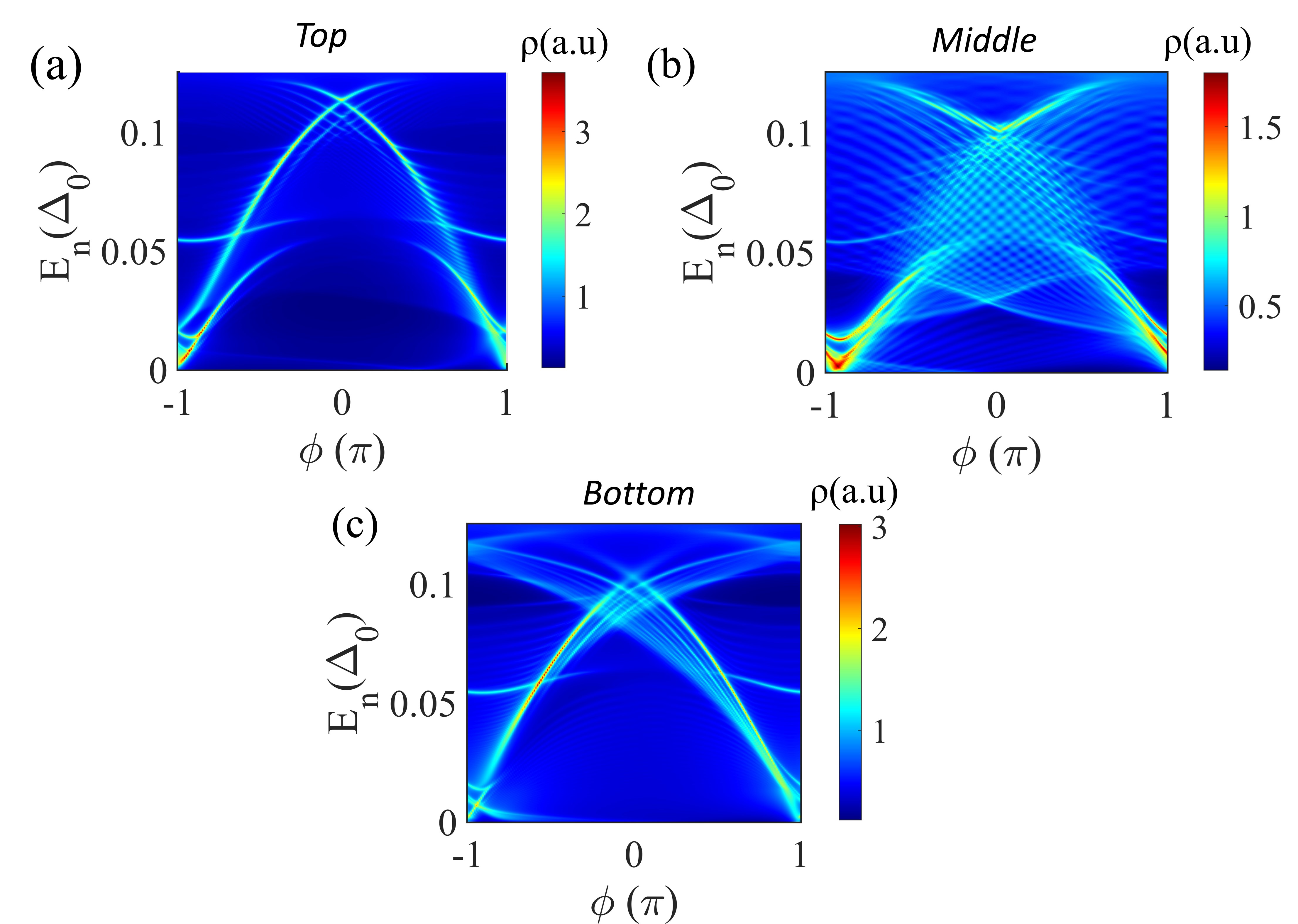}
\caption{LDOS at the (a) top edge, (b) middle, and (c) bottom edge of the channel with $d+id^\prime$ pairing in the leads. The peaks of  LDOS correspond to the ABS energies. Most of the bound states are present at both the top and bottom. However,  a few bound states only appear at one end. The bound states in the middle of the channel are significantly different from the two edges.  $\mu = 0.1\:t$, $\Delta_0 = 0.01\:t$, $\Delta^\prime_0 = 0.001\:t$, $\theta_L = -\theta_R = 0.15\:\pi$, $L_y = 40\:\xi$  and $W_J = \xi/2$. }
    \label{fig:ABS}
\end{figure}

\begin{figure}
    \centering
    \includegraphics[width=\columnwidth]{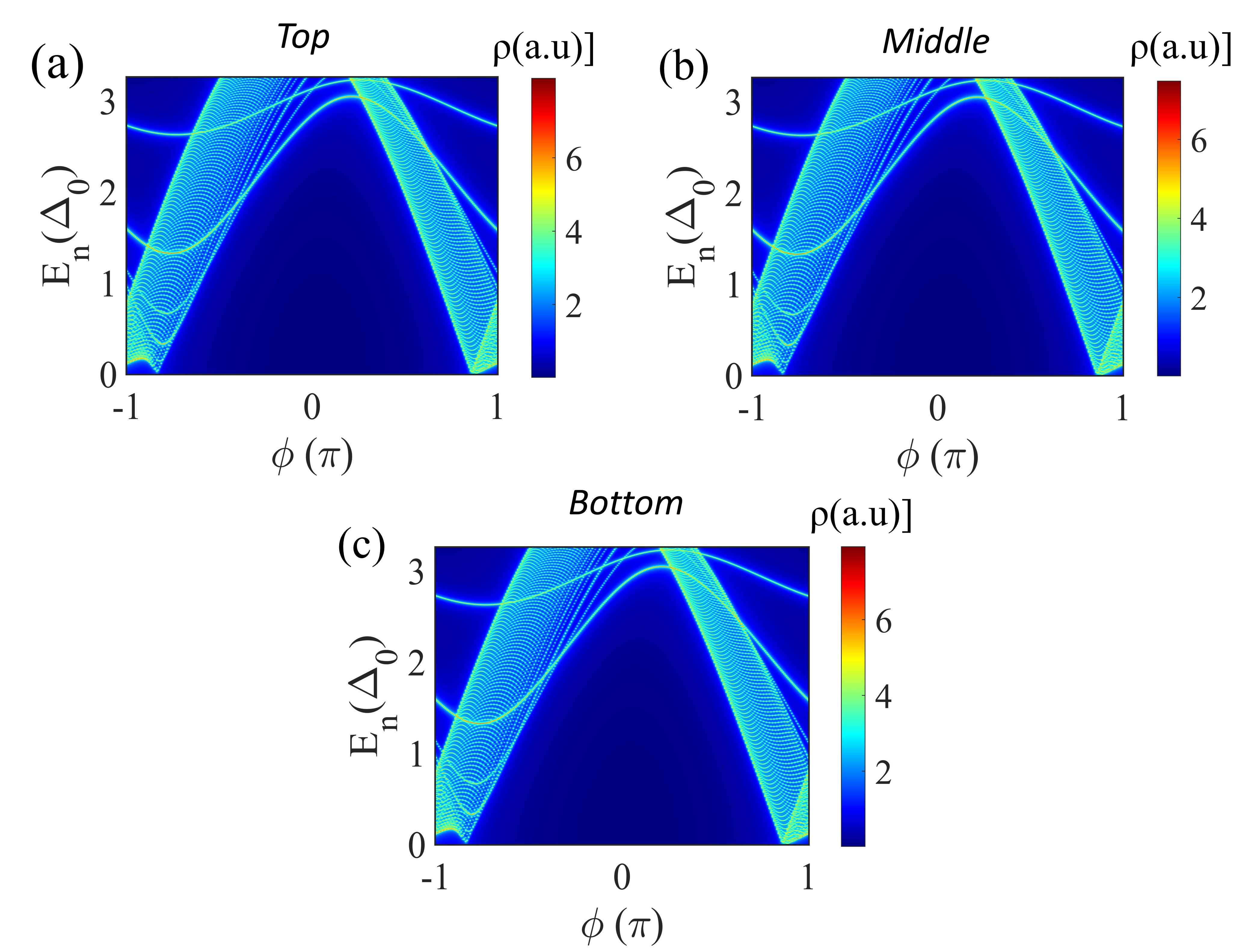}
\caption{LDOS at the (a) top edge, (b) middle, and (c) bottom edge of the channel with $d+is$ pairing in the leads. Parameters used: $\mu = 0.1\:t$, $\Delta_0 = 0.01\:t$, $\Delta^\prime_0 = 0.001\:t$, $\theta_L =0$, $ \theta_R = 0.5\:\pi$, $L_y = 20\:\xi$   and $W_J = \xi/2$.  }
    \label{fig:ABS2}
\end{figure}

Next, we look at the LDOS to identify the behavior of ABS in the vicinity of edges. Fig.~\ref{fig:ABS}(a) shows the LDOS at  the top edge of the channel for a system with leads featuring $d+id^\prime$ pairing. We observe a clear asymmetry as $\phi$ is changed which signifies the nonreciprocity of the current density. In  Fig.~\ref{fig:ABS}(b), we show the LDOS calculated in the middle of the channel, which also reveals the asymmetry and nonreciprocity. However, the pattern of the LDOS is clearly different compared to the LDOS at the top edge. This agrees with the behavior of the current density in Fig.~\ref{fig:currentdensity} attributed to the presence of edge states. The LDOS at the bottom edge in Fig.~\ref{fig:ABS}(c) resembles the LDOS of the top edge. Above behavior is in contrast with Fig.~\ref{fig:ABS2} calculated for $d+is$ pairing where we observe similar patterns for the top, bottom, and middle parts, and no signatures of edge states.

\begin{figure}
    \centering
    \includegraphics[width=1\columnwidth]{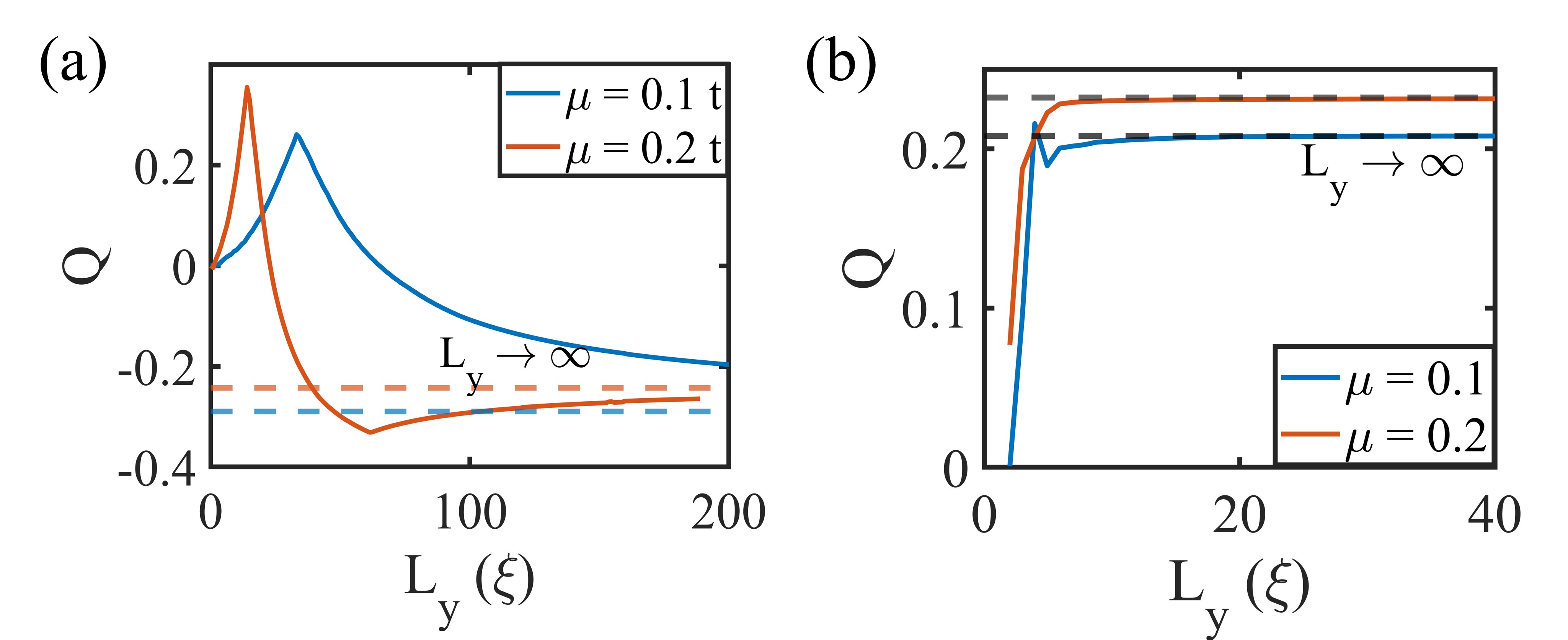}
    \caption{Q is plotted as $L_y$ is increased for $d+id$ (a) and $d+is^\prime$ (b) pairing in the leads. The peak in Q for $d+id^\prime$ pairing corresponds to the edge-dominated transport. Two sets of parameters used: $\mu = 0.1\:t$, $\Delta_0 = 0.1\:t$, $\Delta^\prime_0 = 0.01\:t$,  and $\mu = 0.2\:t$, $\Delta_0 = 0.1\:t$, $\Delta^\prime_0 = 0.01\:t$. $W_J = \xi/2$ for both sets. }
    \label{fig:QvsLy}
\end{figure}

To separate the edge contributions from the bulk contributions, we look at the behavior of $Q$ as the transverse size $L_y$ is increased. 
Note that to make the calculation feasible, we had to choose unrealistically large value of the pairing potential $\Delta_0$. Figure~\ref{fig:QvsLy}(a) shows that in case of $d+id^\prime$ pairing, a JJ with a relatively narrow channel ($L_y \sim \xi$) exhibits interplay of the edge and bulk contributions where $Q$ reaches a maximum. As $L_y$ is increased further, the bulk current contribution becomes dominant, which results in the sign reversal of $Q$. We eventually get close to the asymptotic value corresponding to $L_y \rightarrow \infty$ where $Q$ is calculated using the Fourier transform approach that ignores the edge contributions. In comparison, for a similar calculation with leads featuring the $d+is$ pairing, the asymptotic limit of the channel width is reached very quickly ($L_y \approx 2\xi$), as shown in Fig.~\ref{fig:QvsLy}(b).

Similar to the JJ considered in the previous section, the systems featuring $d+id^\prime$ pairing can be tuned by a gate potential in the normal region. This is demonstrated in Fig.~\ref{fig:Q_mut} where $Q$ is calculated as a function of the gate potential $V_G$. The strong dependence on the gate voltage is observed for $\theta_L$ and $\theta_R$ that are in the vicinity of points where $Q$ changes sign in Fig.~\ref{fig:Q_phase}(a), as can be seen by comparing different curves in Figs.~\ref{fig:Q_mut}(a) and (b). We can achieve tunability and even the sign change, thus replicating the transistor functionality.  

 \begin{figure}
    \centering\includegraphics[width=\columnwidth]{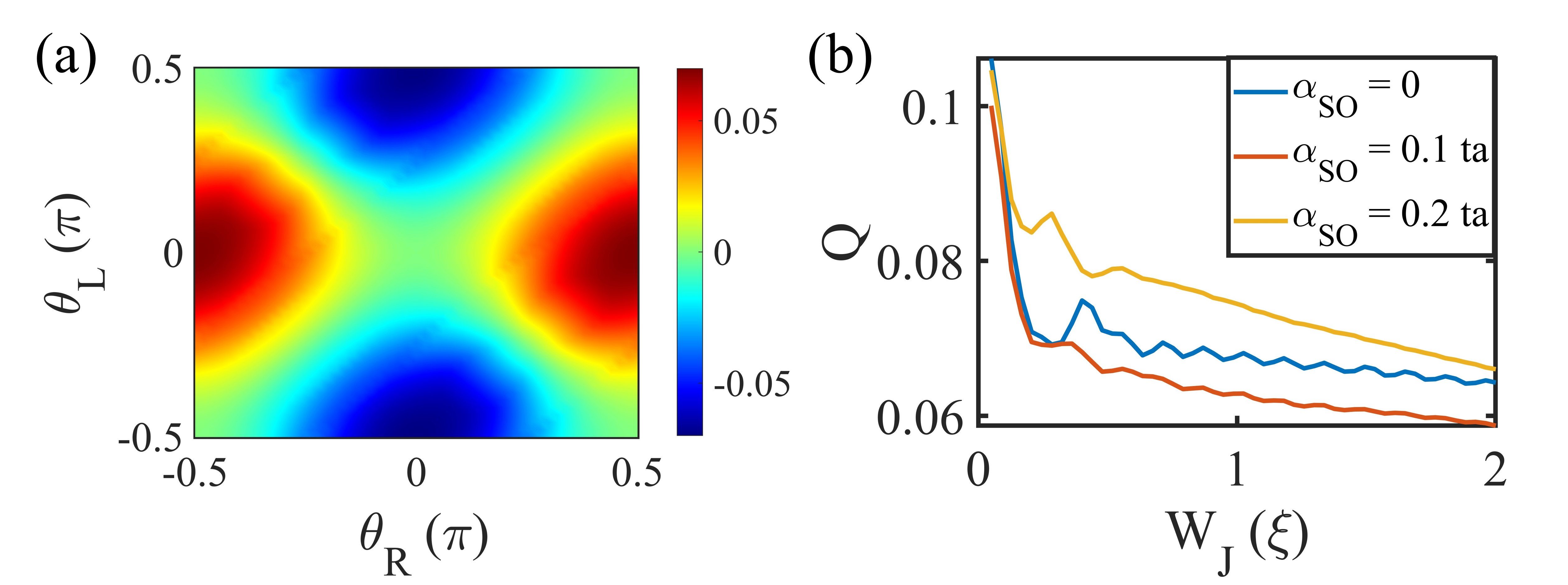}
\caption{(a) Phase diagram of Q as $\theta_L,\theta_R$ are varied in a $d+is|$N$|d+is$ JJ. Along the $\theta_R=\pm \theta_L$ lines, Q is zero. A sign change also occurs at these lines. (b) Plot of Q for various $\alpha_{SO}$ as $W_J$ is increased with $\theta_L=0$, $\theta_R=0.5\:\pi$. Parameters used: $\mu = 0.1\:t$, $\Delta_0 = 0.01\:t$, $\Delta^\prime_0 = 0.001\:t$, $k_BT = 0.01\:\Delta_0$, and $W_J = \xi/2$ for (a). The calculations are performed using a Fourier transform along the y-axis}
    \label{fig:Q_d+is}
\end{figure}

\section{Leads with $d \pm is$ pairing}
\begin{figure}
    \centering\includegraphics[width=\columnwidth]{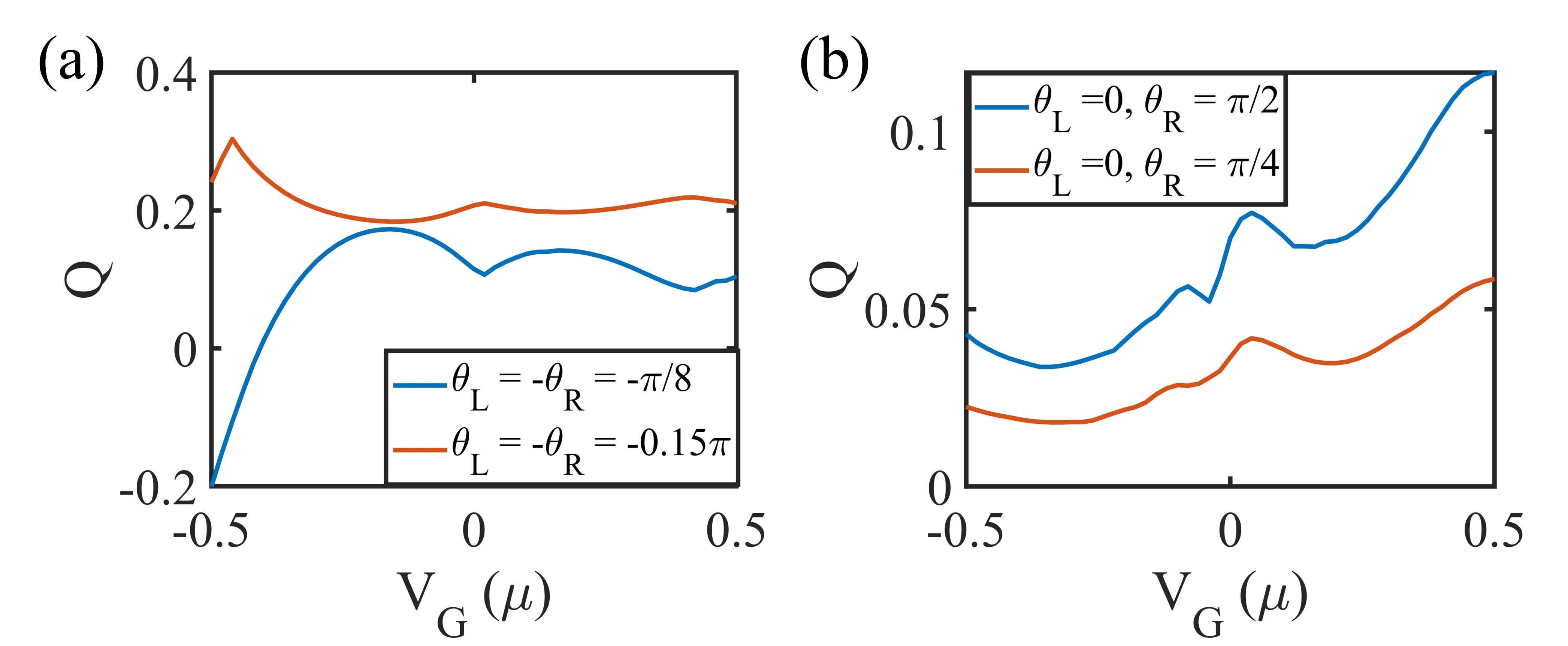}
\caption{The plot of Q as $V_G$ is varied for $d+id^\prime$ (a) and $d+is$ (b) pairing in the leads. Parameters used: $\mu = 0.1\:t$, $\Delta_0 = 0.01\:t$, $\Delta^\prime_0 = 0.001\:t$, $k_BT = 0.01\:\Delta_0$, $W_G = \xi/4 $, and $W_J = \xi/2$.   }
    \label{fig:Q_mut}
\end{figure}

Finally, we consider a ballistic JJ with superconducting leads featuring the $d\pm is$ pairing.
Figure~\ref{fig:Q_d+is} (a) shows the quality factor $Q$ as a function of the orientation angles $\theta_L$ and $\theta_R$. 
For a $d+is|$N$|d+is$ JJ, we observe in Fig.~\ref{fig:Q_d+is}(a) that there is no JDE for orientations described by $\theta_L=\theta_R$ or $\theta_L=-\theta_R$, which is consistent with our symmetry analysis. The vanishing JDE for orientations described by $\theta_L=-\theta_R$ is in contrast with the case of $d+id^\prime$ pairing in Fig.~\ref{fig:Q_phase}(a), and can be used as a signature to distinguish between the two types of pairings experimentally. The quality factror a for $d-is|$N$|d+is$ JJ can be obtained from Fig.~\ref{fig:Q_d+is}(a) by a transformation $\theta_L \rightarrow \theta_L + \pi/2$ or $\theta_R \rightarrow \theta_R + \pi/2$. This follows from Eq.~\eqref{eq:dis} as the transformation $\theta \rightarrow \theta + \pi/2$ changes $d-is$ to $-d-is$. In Fig.~\ref{fig:Q_phase}(b), we study the variation of $Q$ with the junction length $W_J$ and find that the quality factor decreases by a factor of two for a longer junction. Furthermore, the presence of SOC only leads to minor changes. In Figs~\ref{fig:currentdensity}(f, g), we examine the current distribution along the transverse direction for a JJ of finite width $L_y$. In Fig.~\ref{fig:currentdensity}(f) for a $d+is|$N$|d+is$ JJ, we observe a uniform distribution of current along the $y$ direction for orientation angles $\theta_L=0$ and $\theta_R=0.5\pi$. In Fig.~\ref{fig:currentdensity}(g) for orientation angles $\theta_L=\theta_R=0.15\pi$, we observe only a minor difference between the edge and the bulk and a vanishing diode effect. To further identify the differences between $d+id^\prime$ and $d+is$ pairings, we study the width dependence of JDE in Fig.~\ref{fig:QvsLy}. For the $d+is$ case in Fig.~\ref{fig:QvsLy}(b), we observe that JDE quickly reaches the asymptotic value, which is consistent with the absence of edge contributions for $d\pm is$ pairing.
In Fig.~\ref{fig:Q_mut}(b), we study how a JJ with $d+is$ pairing can be tuned by a gate potential in the normal region. As in the previously considered cases, we observe that the qualuity factor $Q$ can be tuned by the gate voltage $V_G$, especially for orientation angles $\theta_L$ and $\theta_R$ in Fig.~\ref{fig:Q_d+is}(b) corresponding to the sign change of $Q$.  
\section{Conclusions} 
In this work, we have studied JDE in S$|$N$|$S Josephson junction formed using 2DEG and superconductors with $d$, $d+id^\prime$, and $d+is$ superconducting pairings. We have shown that the quality factor and its sign 
$Q$ can be substantially tuned by the external magnetic field, gate voltage, and the length of the junction for all three types of pairings.
We have demonstrated that due to the intrinsic TRS breaking in $d+is$ and $d+id^\prime$ superconductors, we can realize field-free JDE, where even a relatively small TRS-breaking superconducting pairing component can result in a large JDE.  In particular, by breaking certain rotational symmetries a large JDE is possible in the absence of the Zeeman field and SOC. By calculating LDOS and the current distribution inside the normal region, we have also studied the role of edge states associated with $d+id^\prime$ pairing. We see clear signatures of the edge states in systems with $d+id^\prime$ pairing where the edge currents can have a significant contributions to JDE. We find that JDE vanishes in systems with $d+is$ and $d+id^\prime$ pairings under different symmetry constraints. This leads us to the conclusion that JDE can be used to differentiate between different superconducting pairings. Our proposal can be useful for superconductor rectification and realizations of transistor-like functionalities, e.g., also using magnetic textures~\cite{PhysRevB.108.174516,Gngrd2022}, in superconducting devices operating at low temperatures.

\acknowledgements
This work was supported by the National Science Foundation/EPSCoR RII Track-1: Emergent Quantum Materials and Technologies (EQUATE), Award OIA-2044049. 

\bibliography{main}
\bibliographystyle{apsrev4-2}
\end{document}